\documentclass{nature}
\usepackage{graphicx}
\usepackage{amsmath,amssymb}
\usepackage[utf8]{inputenc}
\usepackage{bibentry}
\usepackage{hyperref}
\usepackage{color}
\hypersetup{colorlinks=true,citecolor={blue},linkcolor={blue},urlcolor={blue}}
\usepackage{units}
\usepackage[english]{babel}
\usepackage{dcolumn}
\usepackage{bm}
\usepackage{xfrac}
\usepackage[normalem]{ulem}
\usepackage{calrsfs}
\usepackage{nicefrac}

\title{Giant increase of temporal coherence in optically trapped polariton condensate}

\author{A. Askitopoulos$^{1}$, L. Pickup$^{2}$, S. Alyatkin$^1$, A. Zasedatelev$^{1}$, K.G. Lagoudakis$^3$, W. Langbein$^4$ \& P.G. Lagoudakis$^{1,2}$}

\makeatletter
\let\saved@includegraphics\includegraphics
\AtBeginDocument{\let\includegraphics\saved@includegraphics}
\renewenvironment*{figure}{\@float{figure}}{\end@float}
\makeatother
\begin{document}
\renewcommand{\thefootnote}{\alph{footnote}}	
\newcommand{\Fig}[1]{Fig.\,\ref{#1}}
\newcommand{\fig}[1]{fig.\,\ref{#1}}
\maketitle

\begin{affiliations}
 \item Skolkovo Institute of Science and Technology, Novaya St., 100, Skolkovo 143025, Russian Federation
 \item Department of Physics and Astronomy, University of Southampton, Southampton, SO17 1BJ, United Kingdom.
 \item Department of Physics, University of Strathclyde, Glasgow, G40NG, United Kingdom
 \item School of Physics and Astronomy, Cardiff University, Cardiff, CF24 3AA, United Kingdom
\end{affiliations}
\begin{abstract}
Coherent bosonic ensembles offer the promise of harnessing quantum effects in photonic and quantum circuits. In the dynamic equilibrium regime, the application of polariton condensates is hindered by exciton-polariton scattering induced decoherence in the presence of a dark exciton reservoir. By spatially separating the condensate from the reservoir, we drive the system into the weak interaction regime, where the ensemble coherence time exceeds the individual particle lifetime by nearly three orders of magnitude. The observed nanosecond coherence provides an upper limit for polariton self-interactions. In contrast to conventional photon lasers, we observed an increased contribution from the super-Poissonian component of the condensate to the overall particle number fluctuations. Coupled with the recent emergence of a quantum regime in polaritonics, coherence times extended to several nanoseconds favour the realisation of quantum information protocols.
\end{abstract}

\section*{Introduction}
Polariton condensates in semiconductor microcavities are promising systems not only for the study of Bose-Einstein Condensation (BEC), but also as a new platform for the implementation of photonic and quantum circuits~\cite{fraser_physics_2016,espinosa-ortega_complete_2013}. Polaritons are a superposition of photons and excitons~\cite{deng_exciton-polariton_2010} that due to their light effective-mass ($\sim 10^{-5}m_e$ ) can undergo Bose-Einstein condensation even at room temperature ~\cite{christopoulos_room-temperature_2007,plumhof_room-temperature_2014,daskalakis_nonlinear_2014}, alleviating technological and experimental difficulties associated with low temperature systems like cold atoms and indirect excitons. Thus polariton condensates offer a semiconductor technology platform, where coherent photonic and quantum circuits can be implemented in close analogy to existing proposals and implementations in atomic gases~\cite{amico_focus_2017}. To this end, the interaction of spatially separated polariton condensates has been studied both in etched structures~\cite{kim_dynamical_2011,jacqmin_direct_2014,sala_spin-orbit_2015,suchomel_platform_2018} and in optically molded potentials and lattices~\cite{berloff_realizing_2017,ohadi_spin_2017,ohadi_nontrivial_2016,alyatkin_optical_2019}. Whereas the particle lifetime of these coherent ensembles is limited to a few picoseconds, it is the condensate coherence time that effectively dictates the scalability and versatility of the system, limiting the extent into which condensates maintain their macroscopic phase coherence before domain fragmentation~\cite{krizhanovskii_coexisting_2009}.

In polaritonics, the transition to the condensed phase is evidenced by the build-up of off-diagonal long-range order,  a sharp decrease of the emission linewidth, a non-linear increase of the emission intensity, and a blueshift of the emission energy associated with the repulsive nature of interparticle interactions~\cite{kasprzak_bose-einstein_2006}.
The primary mechanism of decoherence in polariton Bose gases is pair-particle scattering, which are further amplified by particle number fluctuations in the condensed state~\cite{laussy_coherence_2004}. In the limit of non-interacting particles, an upper bound for the coherence time comes from the Schawlow-Townes limit, similar to conventional lasers~\cite{schawlow_infrared_1958}. Under non-stationary excitation conditions, previous studies on particle number fluctuations have shown that the second order auto-correlation function, ($g^{(2)}(\tau)$), follows a quasi-thermal distribution at threshold, whilst for increasing polariton density recovers Poissonian statistics~\cite{tempel_characterization_2012,whittaker_polariton_2017,klaas_evolution_2018}. The transient dynamics of $g^{(2)}(0)$ indicate that above threshold the system rapidly transitions from a thermal  to a coherent state within a few picoseconds~\cite{adiyatullin_temporally_2015}.

In the dynamic equilibrium regime, where the condensate depletion is continuously replenished by a continuous wave (CW) non-resonant optical pump, there are only a few reports of photon bunching~\cite{love_intrinsic_2008,kasprzak_second-order_2008,kim_coherent_2016}, but a purely thermal distribution remains elusive both above and below condensation threshold.

Above threshold and in the limit of second order coherence, the most prominent dephasing mechanism is driven by particle interactions within the condensate. Under non-resonant steady-state excitation, the presence of an incoherent exciton reservoir is the dominant source of dephasing as condensate-reservoir interactions ($U_{xp}$) are stronger than pair-polariton interactions ($U_{pp}$) in the condensate. In the optical parametric oscillator (OPO) configuration, wherein polaritons are injected resonantly onto the polariton dispersion minimising the presence of the exciton reservoir, coherence times of up to \unit[500]{ps} were reported\cite{krizhanovskii_dominant_2006}. While the polariton interaction strength has been previously measured and calculated with a variety of experimental and analytical methods, the reported values cover a range of more than two orders of magnitude ~\cite{tassone_exciton-exciton_1999,vladimirova_polarization_2009,sun_direct_2017,tsintzos_electrical_2018,estrecho_direct_2019}. Since the coherence time of the condensate is directly affected by pair-polariton interactions as well as polariton interactions with the reservoir, it imposes an upper limit for pair-polariton interaction strength, when the effects of  particle number fluctuations are properly quantified.

In this letter, we report several nanoseconds polariton condensate coherence time in the dynamic equilibrium regime by eliminating the contribution of an exciton reservoir, utilising a combination of scan-less variable delay interferometric measurements and active density stabilization of the condensate. We observe that for condensates formed in an optical trap configuration~\cite{askitopoulos_polariton_2013}, where polariton interactions with the reservoir are eliminated, the coherence time of the condensate increases by almost a factor of 100 in comparison with a polariton condensates overlapping with the exciton reservoir. Second order coherence measurements show that although the condensate density fluctuations above threshold decrease, and with increasing density converge towards Poissonian values, a super-Poissonian noise term persists even above threshold, contributing to decoherence. By actively tuning the condensate density, we observe that the coherence time is proportionally affected. We evaluate polariton and exciton interaction strengths ($U_{pp},U_{xx}$ respectively) by determining the mean number of particles in the condensate and obtain consistent values with previously reported theoretical estimates~\cite{tassone_exciton-exciton_1999,ciuti_role_1998}. These results are juxtaposed to the regime wherein the reservoir spatially overlaps with the condensate, where we observe that the coherence time is strongly reduced and the timescale and amplitude of number fluctuations in the system depend on the condensate size. Finally, we present a triple-logarithmic numerical study of the coherence time of the system on pair-polariton interaction strength, particle number fluctuations and condensate densities that coupled with our experimental findings, allows us to distinguish polariton condensation from conventional laser phase transitions.

\section*{First order Coherence}
\begin{figure}
	\center	
	\includegraphics[scale=0.45]{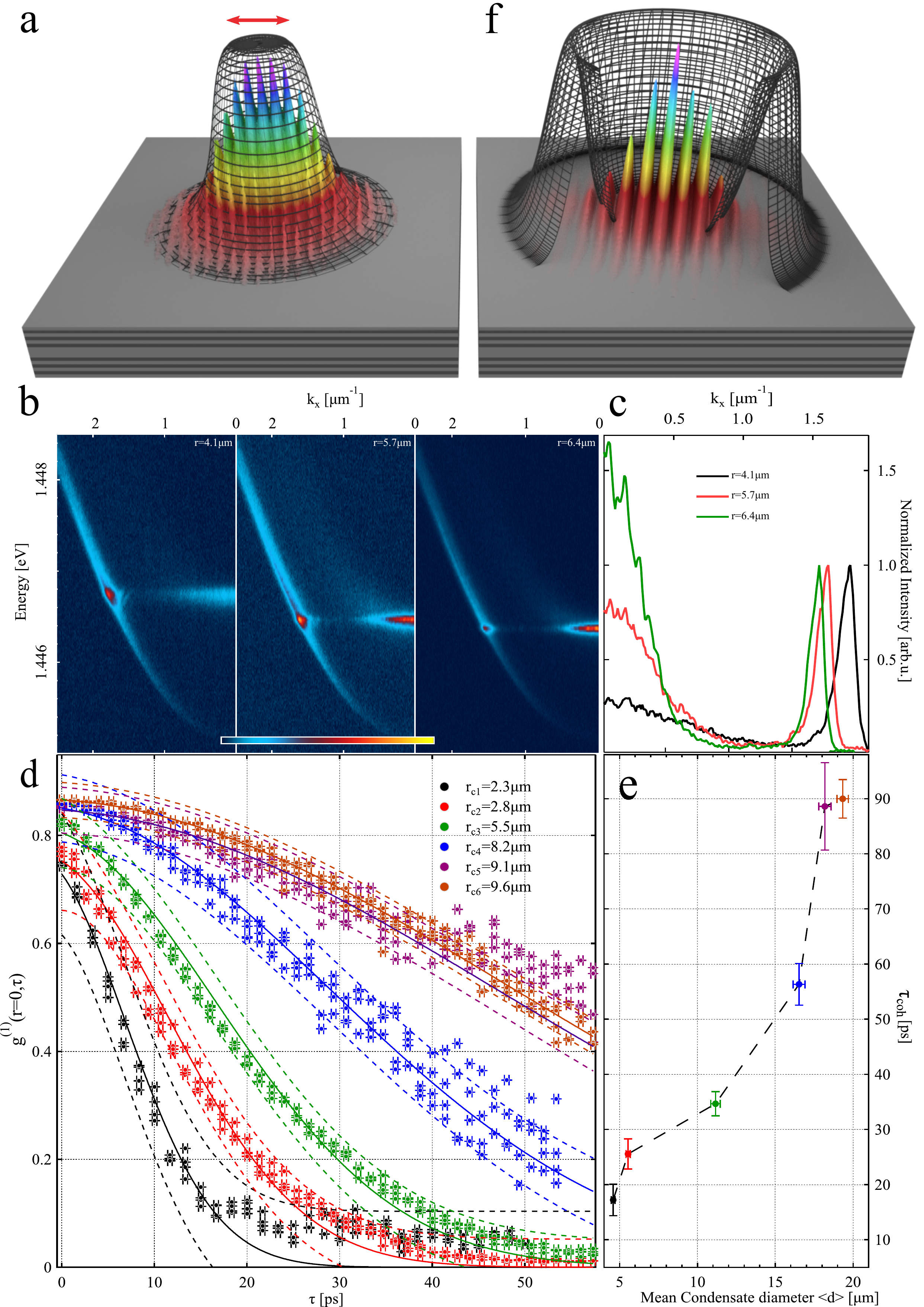}
	\centering	
	\caption{a. Schematic representation of the top hat pumping scheme. b. Energy dispersion vs $k_x$ for 3 top hat excitation sizes at $\approx P_{th}$ and extracted profiles c. d. First order coherence function $g^{(1)}(r=0,\tau)$, for varying condensate radii $r_c$, solid lines are Gaussian fits to the data and dashed lines 95\% prediction bands. e. Extracted coherence time $t_{coh}$ vs mean condensate diameter <d>. Schematic representation of optically trapped condensate f.}
	\label{fig1}
\end{figure}

Previous reports on polariton condensates with a Gaussian non-resonant CW optical pump demonstrated phase coherence times of the order of \unit[100]{ps}~\cite{kasprzak_second-order_2008,kim_coherent_2016}. The theoretical framework utilised in these studies, predicts a power law scaling of coherence time, $t_{coh}$, with condensate area, $A_c$, ($t_{coh}\propto\sqrt{A_c}$) assuming a size independent threshold condensate density~\cite{whittaker_coherence_2009}. 
The most evident spectral difference between condensates of different sizes is the in-plane wave-vector, where condensation occurs~\cite{wouters_spatial_2008}. Condensation in a non-zero wave-vector is a direct consequence of the repulsive interactions with the localized exciton reservoir. The latter forms a potential hill co-localised with the pump from which polaritons are  expelled by converting potential energy to kinetic energy ~\cite{christmann_polariton_2012,askitopoulos_robust_2015}. To reduce the distribution of wavevectors acquired by the condensate, we inject a uniform exciton reservoir density by using a non-resonant CW top hat (TH) excitation profile (Fig.\ref{fig1}a) and we study the phase coherence time of a condensate for different diameters in the dynamic equilibrium regime. Fig.\ref{fig1}b shows the PL of the polariton dispersion for four different top hat diameters at condensation threshold. We integrate  the polariton emission over twice the full width half maximum (FWHM) of the polariton condensate linewidth and extract the relative wave-vector distribution, shown in Fig.\ref{fig1}c. Evidently, the broader the diameter of the top hat excitation, the higher the relative population of polaritons around zero wavevector.

We use a Michelson-Morley interferometer and measure the first order coherence ($g^{(1)}(\tau)$) for varying delay times, (see Supplementary Information SI and video (SV1)). The coherence time is obtained by fitting a Gaussian function to the data and evaluating the integral~\cite{loudon_quantum_2000}:
\begin{equation}
\tau_{coh}=\int_{-\infty}^{\infty} \left| g^{(1)}(\tau)\right|^2d\tau
\label{eqtc}
\end{equation}
For a tightly focused top hat excitation (condensate radius ($r_{c1}=\unit[2.3]{\mu m}$), extracted at the half-Width at half maxima) and excitation power $\approx$30\% above threshold ($P=1.3P_{th}$), we observe a fast decay ($\tau_{coh}\approx17 ps$) of the ensemble phase coherence,  Fig.\ref{fig1}d, comparable to the cavity lifetime ($\unit[7]{ps}$). For larger condensate sizes while keeping the power relative to threshold constant at $P=1.3P_{th}$, the initial coherence of the system ($g^{(1)}(0)$) increases and the decay is prolonged up to $\unit[90]{ps}$. Fig.\ref{fig1}e shows the extracted coherence time as a function of the mean condensate diameter ($\langle FWHM\rangle$). While this observation is in line with the predictions of the theoretical framework for phase coherence of polariton condensates~\cite{whittaker_coherence_2009}, it is noteworthy that the increase of coherence time correlates with the decrease in condensation wave-vector. This in turn, relates to the excitation density dependence of the potential formed by the exciton reservoir, that is the dominant form of de-coherence in the system. Indeed when the pump size is of the order of the polariton de-Broglie wavelength ($\lambda_{dB}=\sqrt{\nicefrac{2\pi\hbar^2}{m_pk_BT}}=\unit[3.3]{\mu m}$), polaritons quickly escape the pumping region and higher reservoir densities are needed to drive the system to the condensation regime.

\begin{figure}
	\center	
	\includegraphics[scale=0.55]{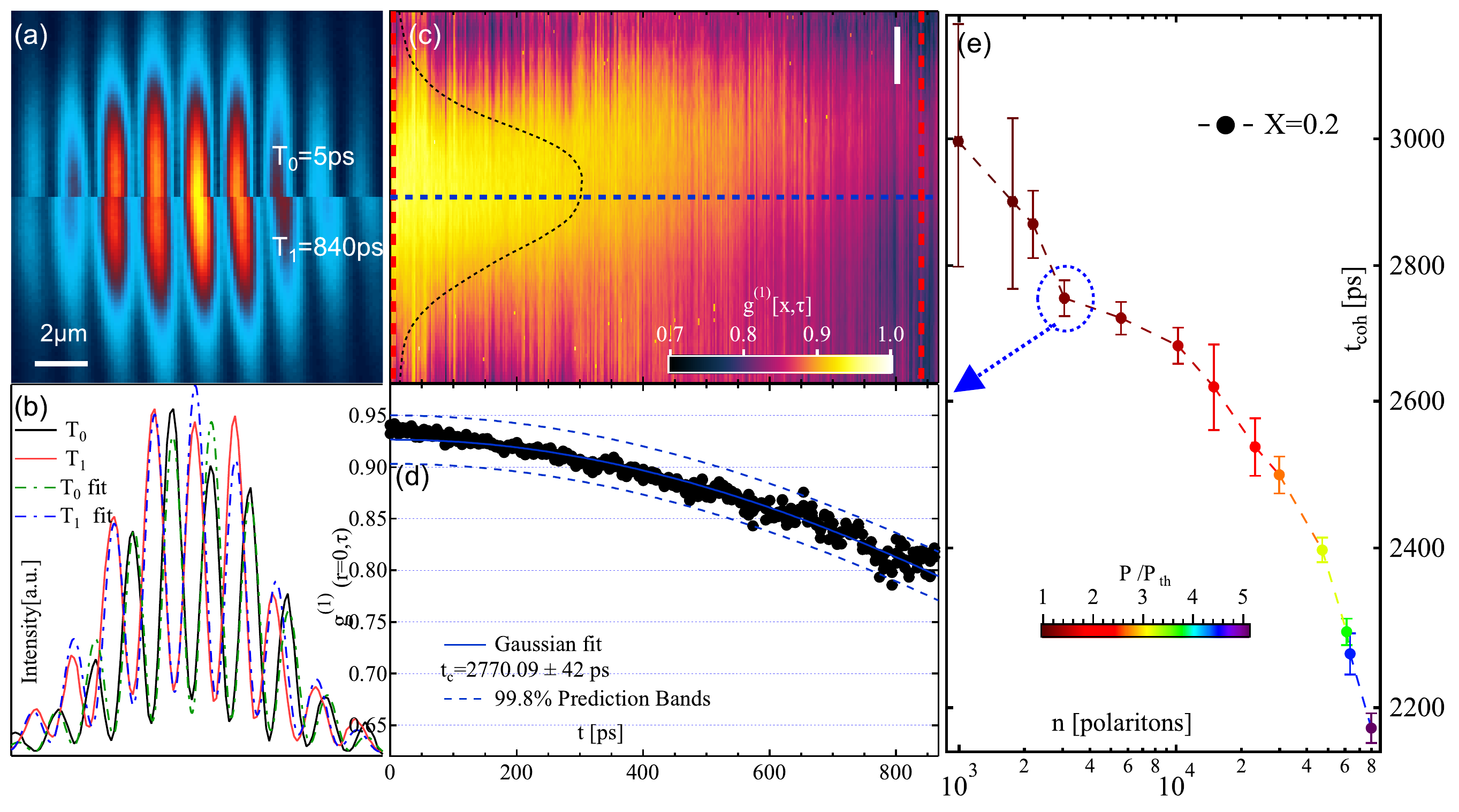}
	\centering
	\vspace{-10pt}
	\caption{First order coherence measurements $g^{(1)}[x,t]$ for a polariton condensate in the optical trap configuration. (a) Real space interferograms and corresponding profiles (b) at $T_0=\unit[5]{ps}$ and at $T_1=\unit[840]{ps}$ showing the degradation of the interference fringes. (c) Extracted $g^{(1)}[x,\tau]$, the black dashed line marks the condensate real space intensity and the red dashed lines mark the delay times depicted in (a) and (b). (d) $g^{(1)}[x=0,t]$, black circles, with Gaussian fit, blue line, and $99.8\%$ confidence prediction bands. The points are averaged within 5 pixels around the dashed blue line of (c). (e) Extracted coherence times as a function of polariton density, the colour-scale denotes the power relative to threshold. }
	\label{fig2}
\end{figure}

It is thus meaningful to examine if the condensate coherence time can be extended by further minimizing reservoir interactions.
We experimentally realize a condensate in the optical trap (OT) configuration as sketched in Fig.\ref{fig1}f, with a similar excitation power above threshold ($P=1.3P_{th}$). The recorded interferograms remarkably show a very limited suppression of coherence even for a delay of \unit[840]{ps} (Fig.\ref{fig2}a,b). Indeed, the extracted $g^{(1)}[x,t]$ shows a degree of first order coherence in the central region (away from the reservoir) of the condensate of $\sim0.8$ even at the end of our delay line (Fig.\ref{fig2}c,d), in stark contrast to the results obtained with the top hat excitation. Fitting the temporal decay with a Gaussian function and with the use of ~eq.\ref{eqtc} we obtain a coherence time of $\approx \unit[2.7]{ns}$ for the trapped condensate, more than $80$ times longer than a condensate of similar size ($r_{cOT}=3.3\unit{\mu m}$) and TH excitation conditions (Fig.\ref{fig1}d). Although the increase of the coherence time is surprisingly large, this result is indicative of the reduced interactions of the condensate with the reservoir and of the reduced excitation density threshold required for condensation, which further reduces the influence of the reservoir.
In the limit of zero excitonic fraction the coherence of the system should coincide to that of a photon laser. In this regime, the Schawlow-Townes effect leads to a progressive increase of the coherence time with increasing occupancy of the lasing mode. For our matter-wave laser, this effect is counterbalanced by interaction induced dephasing that reduces the coherence time with increasing occupancy ~\cite{kasprzak_bose-einstein_2006,whittaker_coherence_2009,kim_coherent_2016}. Therefore, the occupancy dependence of coherence time provides a hallmark in the coherence features for matter-wave lasers that corroborates the differences between photon and polariton lasers, i.e. lasing in the strong vs the weak coupling regime.

It is therefore meaningful to explore the dependence of the coherence time as a function of condensate density. We thus proceed to record the coherence time of a trapped condensate for increasing excitation density at a polariton exciton fraction of X=0.2, Fig.\ref{fig2}e (see also SV2). We note that for increasing condensate density the coherence time of the system is decreasing, but is still close to 2 orders of magnitude longer than that of a TH condensate of similar size. In the context of established analytical quantum models for matter-wave lasers~\cite{thomsen_atom-laser_2002} that have also been applied to inorganic polariton condensates~\cite{love_intrinsic_2008,whittaker_coherence_2009,kim_coherent_2016}, the temporal decay of the first order coherence is expressed as

\begin{equation}\\\label{eqg1}
\left|g^{(1)}(\tau)\right|= \exp\left(- 4\sigma_{{n}}^2 \left(\bar{\tau}U_{pp}\right)^2\left(\exp \left(-\nicefrac{\tau}{\bar{\tau}} \right) +\nicefrac{\tau}{\bar{\tau}} -1 \right)\right) \\
\times \exp\left( \frac{\sigma_{{n}}^2}{4\bar{n}^2}\left(\exp \left(-\nicefrac{\tau}{\bar{\tau}} \right) -\nicefrac{\tau}{\bar{\tau}} -1 \right)\right)
\end{equation}

where $\bar{\tau}=\nicefrac{\sigma_{{n}}^2}{\bar{n}}  t_c$ , $t_c$ is the polariton lifetime,  $U_{pp}$ is the polariton-polariton interaction, $\bar{n}$ is the average number of particles in the condensate and $\sigma_{{n}}^2$ is the variance in the particle number.The first part of Eq.~\ref{eqg1} quantifies the reduction of coherence by particle interactions that scale  with $\sigma_{{n}}^2$, while the second part is similar to the Schawlow-Townes term for lasers and increases the coherence with increasing mode occupancy. Since these are two counteracting effects, it is instructive to evaluate the parameter space of the coherence time in Eqs~\ref{eqtc} and~\ref{eqg1}. Previous reports~\cite{love_intrinsic_2008,whittaker_coherence_2009,kim_coherent_2016} have used the  measured particle number fluctuations at condensation threshold ($\sigma_{{n}}^2=\sigma_{{n}_{th}}^2+\bar{n}$) in order to evaluate the strength of pair polariton interactions. Figure \ref{fig4}a shows the dependence of coherence time on the pair-polariton interaction strength and condensate population for the range of values reported in the literature. Effectively this approach assumes that above condensation threshold all particles relax to the coherent state through bosonic stimulation (with statistics $\sigma_{{n}}^2=\bar{n}$) and predicts an initial plateau and subsequent increase of the coherence time of the condensate with increasing polariton population. However, in ref.\cite{kim_coherent_2016} above threshold a decrease of the coherence time with density was observed and when this model was applied to extract the polariton interaction strength, it was found to be increasing with density, in contrast to theoretical predictions and experimental observation of exciton saturation~\cite{rochat_excitonic_2000,kirsanske_observation_2016}. In Ref.~\cite{love_intrinsic_2008}  the authors report that the coherence time plateaus for higher condensate densities, however, the system that they study is a multi-mode polariton condensate, while the theory used applies to single mode condensates and does not account for mode competition and multi-mode dynamics. Although it is intuitive to assume $\sigma_{{n}}^2=\sigma_{{n}_{th}}^2+\bar{n}$, nevertheless it fails to reproduce the majority of experimental observations where contrary to the phase diagram of Fig.\ref{fig4}a the coherence time decreases with density. Indeed, to our knowledge the vast majority of published literature in polariton condensates feature an increase of the linewidth ($\Delta$v$=\nicefrac{1}{t_{coh}}$) with density regardless of the interaction strength reported~\cite{kasprzak_bose-einstein_2006,christopoulos_room-temperature_2007,balili_bose-einstein_2007,cilibrizzi_polariton_2014,sun_direct_2017,tempel_characterization_2012,askitopoulos_polariton_2013}. It is therefore necessary to evaluate the application of this assumption by investigating particle number fluctuations at and above threshold.  

\section*{Second order coherence}
The density dependence of the underlying statistics of the system are critical for the correct evaluation of the system properties. Optically created high energy and momentum polaritons relax to the ground state of the system through interactions with the reservoir. When the ground state population reaches unity occupancy values, final state bosonic stimulation sets in and speeds up the build up of the population in the ground state, driving the system to quantum degeneracy. Particles reaching the ground state through spontaneous relaxation processes will have a randomized (chaotic) phase~\cite{laussy_coherence_2004} and obey thermal statistics/noise, whereas those injected through bosonic stimulation will obey Poissonian statistics. When these populations are comparable, as is the case at threshold, we expect the signal to exhibit photon bunching. Utilizing a standard Hanbury Brown Twiss (HBT) correlation setup (see also SI) we measure the normalized second order correlation function for the condensate emission for the TH and OT configurations.
\begin{equation}
	g^{(2)}(t,\tau) = \frac {\langle \hat{\alpha}^{\dagger}(t)\hat{\alpha}^{\dagger}(t+\tau) \hat{\alpha}(t)\hat{\alpha}(t+\tau) \rangle }{\langle\hat{\alpha}^{\dagger}(t)\hat{\alpha}(t)\rangle  \langle\hat{\alpha}^{\dagger}(t+\tau)\hat{\alpha}(t+\tau)\rangle}
\end{equation}

Fixing the excitation density around threshold values ($P=(1\pm 0.014)P_{th}$), we observe a clear photon bunching signal for zero time delay,  $g^{(2)}(0)=1.08$, with an exponential decay time of $\approx \unit[400]{ps}$ for the OT condensate (Fig.\ref{fig3}a). For increasing polariton density, the coherent population dominates and photon bunching is substantially suppressed (Fig.\ref{fig3}b). We observe a similar behaviour for the top hat excitation ($r_c\approxeq 10\mu m$), although the photon bunching at threshold is much less pronounced (Fig.\ref{fig3}c) and barely discernible from the measurement noise. We fit $g^{(2)}(\tau)$ with an exponential decay and extract the relevant decay times and amplitudes of second order coherence (Fig.\ref{fig2}d). For the second order coherence properties of the system according to the previously mentioned quantum model, the comparatively slow decay of particle number fluctuations compared to the polariton lifetime (estimated to be $\approxeq\unit[7]{ps}$) is an aftermath of the gain-loss balance of the system near threshold~\cite{haken_cooperative_1975,whittaker_coherence_2009}. In this framework the second order coherence function is expressed as
\begin{equation}\label{eq1}
	g^{(2)}(\tau,\bar{n})-1=  \frac{n_s}{ \bar{n}^2} \exp\left(-\frac{\bar{n}}{\left(\bar{n}+n_s \right)}\frac{\tau}{t_c}\right)
\end{equation}

while in terms of particle number statistics, $g^{(2)}(0)$ is defined as~\cite{loudon_quantum_2000}
\begin{equation}\label{g2s}
	\begin{split}
		g^{(2)}(0)-1=&\frac{\sigma_{\bar{n}}^2-\bar{n}}{ \bar{n}^2}\\
	\end{split}
\end{equation}

where $n_s$ is a gain saturation term that in standard laser theory is proportional to the optical transition rate $\mathcal{\gamma}$ and to the square of the energy difference between initial and final states $\omega_0$ ($n_s\propto \mathcal{\gamma} \omega_0^2$)~\cite{loudon_quantum_2000}. From equations~\ref{eq1},\ref{g2s}, ${n_s}=\sigma_{\bar{n}}^2-\bar{n}$ and the relative variance ($\sigma_{\bar{n}}^2/\bar{n}^2$) of particles in the condensate can be directly inferred from the measured $g^{(2)}(\tau=0)$. For $\bar{n}>>1$ this corresponds to $\frac{\sigma_{\bar{n}}}{\bar{n}}=0.28$ for the optical trapped case and $\frac{\sigma_{\bar{n}}}{\bar{n}}=0.14$ for the top hat condensate.
In the absence of any confinement, polaritons are ejected from the excitation spot, effectively requiring larger reservoir populations to instigate bosonic stimulation. 

\begin{figure}
	\center
	\includegraphics[scale=0.42]{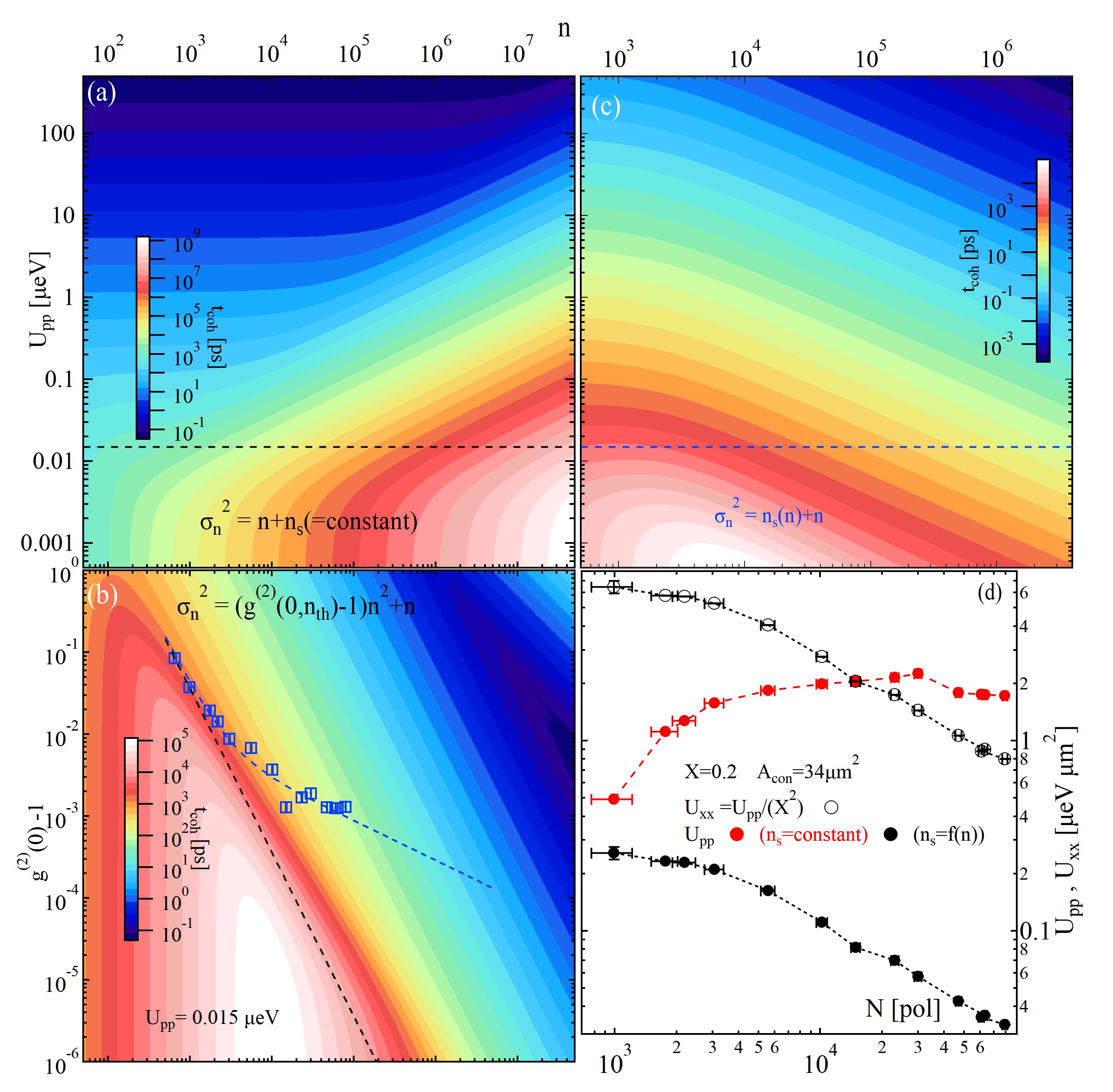}
	\centering
	\caption{ Coherence time parameter space diagrams and extracted interaction strengths. (a) Coherence time for $\sigma_{\bar{n}}^2=n_s(=constant)+n$ for the range of polariton interaction strengths reported in the literature (b). Coherence time for $V_{pp}=\unit[0.015]{\mu eV}$ and $\sigma_{\bar{n}}^2$ defined by $g^{(2)}(0)$. Blue squares denote the values extracted from the experiment and the dashed blue line a fit to the data. The black dashed line marks the  $g^{(2)}(0)$ values for $n_s=constant$ and is equivalent to the black dashed line of (a). (c) Coherence time for $g^{(2)}(0)$ as approximated by the experimental values vs density and $V_{pp}$. The blue dashed line is equivalent in (b) and (c). (d) Polariton-polariton, $U_{pp}$ filled circles,  and exciton-exciton, $U_{xx}(=U_{pp}/X^2)$ open circles,  interaction strength corresponding to the measured coherence times for an exciton fraction of $X=0.2$.}
	\label{fig4}
\end{figure}

\begin{figure}
	\vspace{-10pt}
	\center
	\includegraphics[scale=0.45]{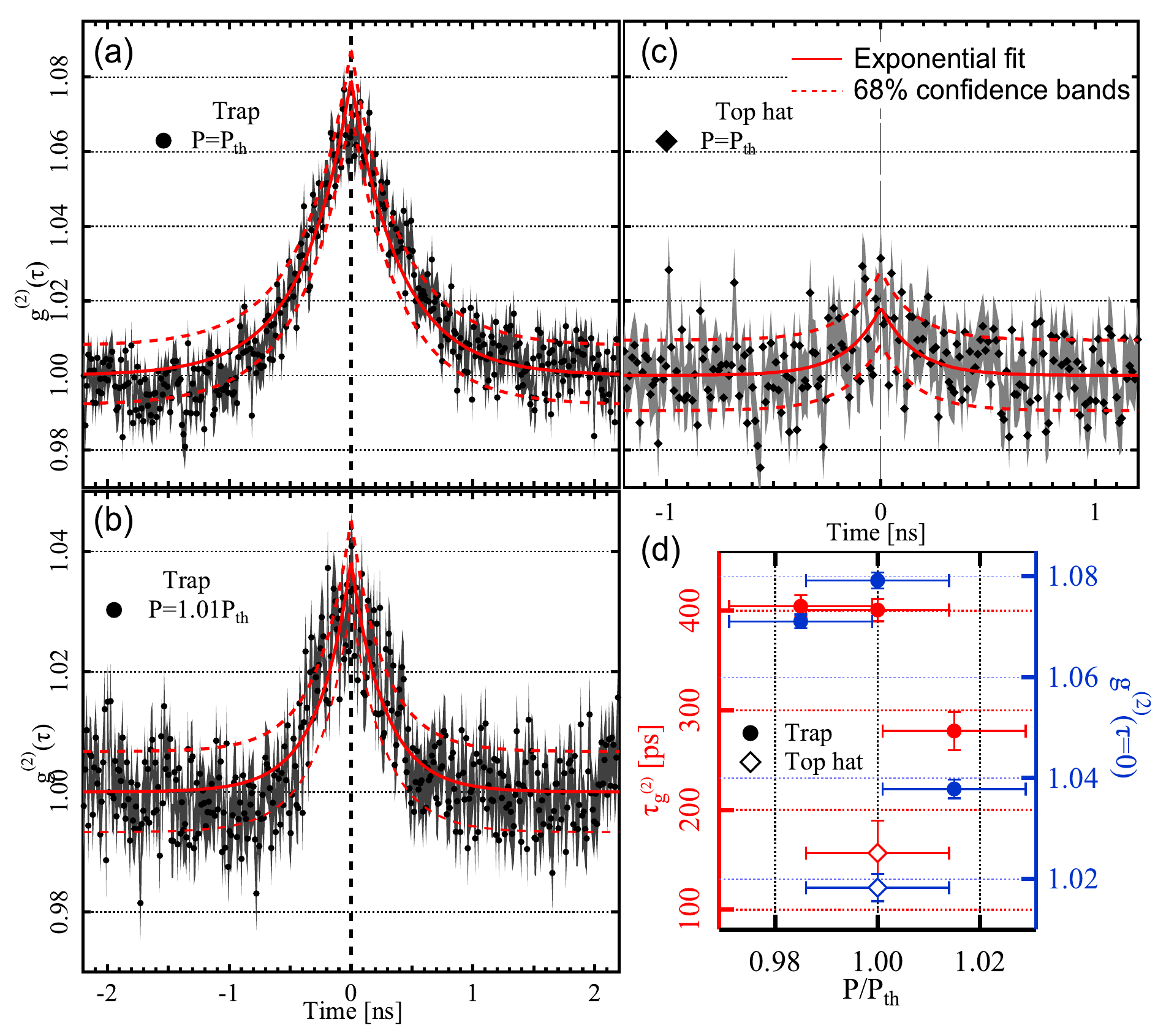}
	\centering
	\caption{Second order correlation measurements $g^{(2)}[\tau]$ for a trapped condensate at different powers above threshold (a) and (b), and for a top hat beam excitation (c). Extracted decay times (red points) and $g^{(2)}(\tau=0)$ amplitude (blue points) for each case (filled circles for the trap and diamonds for the top hat beam)(d). }
	\label{fig3}
	\vspace{-10pt}
\end{figure}

For both cases (TH and OT) increasing the polariton density results in an apparent full inhibition of the bunching signal, within the resolution of our HBT setup, as the denominator in Eq.\ref{eq1} increases. From the measured $g^{(2)}_{th}(\tau=0)$ and the condensate particle number at threshold ($\bar{n}_{th}\approxeq656$, see also SI and using the estimated polariton lifetime of $\approxeq \unit[7]{ps}$~\cite{cilibrizzi_polariton_2014} we get $n_s\approxeq36000$ for the OT. We then use these values to fit our experimental coherence decay curves  with eq.~\ref{eqg1} for different densities with only $U_{pp}$ as a free parameter. Normalizing also with the condensate size ($A_{con}=34\mu m^2$) we get $U_{pp}=\unit[0.5]{\mu eV \mu m^2}$ (red points Fig.\ref{fig4}d) at threshold. However for increasing condensate density this approach gives a non-physical increase of the extracted interaction strength while the opposite is expected as the density approaches the Mott transition, the exciton oscillator strength is quenched and the polariton exciton fraction is reduced (due to blue-shifting of the exciton energy)~\cite{kirsanske_observation_2016}.

At this point it is illuminating to quantify the contribution of super-Poissonian fluctuations and the resolution with which they need to be resolved, we thus examine the system coherence time for $g^{(2)}(\tau=0,n)=constant$, Fig.\ref{fig4}b. While this doesn't correspond to a physical regime, as we have seen that  $g^{(2)}(\tau=0)$ approaches unity as we deviate from threshold, it nevertheless elucidates the pronounced effect that even small deviations from perfect second order coherence can have on the phase coherence time of the system. As shown in Fig.\ref{fig4}b, for high polariton densities even a small super-Poissonian component will drastically amplify dephasing effects from particle interactions.
This in turn indicates the need to estimate $g^{(2)}(\tau=0,n)$ with an error tolerance less than 1\%. As in the HBT measurements the signal to noise ratio (SNR) scales with $\sqrt{t_{meas}}$ it follows that in order to reduce the measurement noise by a factor of 10 we need to increase our integration by a factor of 100 (see also SI). Indeed for many-particle systems, like polariton condensates it is impractical to measure the noise in the single particle regime.
It is however possible to estimate the noise statistics directly from the ensemble of reference measurements (the intensity of a single arm of the interferometer) of $g^{(1)}(\tau)$ due to the statistically important volume of data recorded during the experiment ($250\leq N_{meas}\leq 1500$) and provided that the signal noise dominates the camera noise (see also SI). We find that for $P=P_{th}$ for the optical trapped condensate, $\frac{\sigma_n}{\bar{n}}=0.29$ which is in excellent agreement with the value extracted from the HBT measurement and similar to previously reported values ~\cite{love_intrinsic_2008,kasprzak_second-order_2008}. In Fig.\ref{fig4}b we append the values recovered for increasing density (blue squares) allowing us to now estimate the functional form of $g^{(2)}(0,\bar{n})-1=\nicefrac{n_s(\bar{n})}{\bar{n}^2}$ (blue dashed line Fig.\ref{fig4}b,c) and use it to evaluate $t_{coh}=f(U_{pp},\bar{n})$ (see also SI), shown in Fig.\ref{fig4}c. Using a non-constant, slowly increasing $n_s(\bar{n})$ is enough to recover the experimentally expected dependence of the phase-coherence of the condensate. As the polariton density is known, the only free parameter now remaining is $U_{pp}$ which is subsequently extracted, normalized with the condensate area ($A_{con}$) and shown in Fig.\ref{fig4}d along with the corresponding exciton-exciton ($U_{xx}=U_{pp}/X^2$) interaction strength. 
\section*{Discussion}
The recovered interaction strength around threshold is in excellent agreement with theoretical expectations\cite{ciuti_role_1998,tassone_exciton-exciton_1999} as well as a number of previous experimental observations\cite{vladimirova_polarization_2009,tsintzos_electrical_2018,estrecho_measurement_2018}. Indeed, for the larger interaction strengths mentioned in the literature~\cite{sun_direct_2017}, even under perfect second order coherence the phase coherence time of the system would be well bellow \unit[100]{ps} as can be deduced from \Fig{fig4}a. The observed decrease of interaction strength is expected due to the saturation of the polariton oscillator strength~\cite{deng_exciton-polariton_2010}, as well as a decrease of the exciton interaction strength due to its binding energy being quenched as the system approaches the Mott transition~\cite{ciuti_role_1998,rochat_excitonic_2000,kirsanske_observation_2016}. It is worth pointing out that we also notice a $30\%$ increase of the condensate size from threshold to the highest density measurement, which reflects the increase of the total interactions in the condensate. Our results elucidate than in the context of conventional laser theory, a polariton condensate phase transition can be effectively described as a lasing transition with density dependent gain saturation. It is straightforward to see that the gain saturation changes due to the fact that polaritons are interacting particles and therefore the energy levels that dictate the values for $n_s$ change with density.

\section*{Conclusions}

In conclusion we have investigated the first and second order coherence of polariton condensates and their dependence on condensate size and density in the presence and absence of interactions with the reservoir. We demonstrate that the coherence time of the condensate can extend to the nanosecond regime by quenching the exciton-polariton interaction in the OT and by investigating the parameter space of polariton coherence time we identify the system requirements for obtaining above nanosecond coherence times. Additionally, we demonstrate that when particle number fluctuations are properly accounted for in the system, the recovered polariton interaction strength exhibits the expected values and behaviour with density. Moreover, we have shown that super-Poissonian particle number fluctuations in a polariton condensate are not saturated above threshold, but continue to increase in contrast to lasing from conventional population inversion. The revealed long coherence time regime, enables the study of coherent phenomena, like Bloch oscillations, that have long been studied in atomic systems, as well as enabling robust coherent control in the polariton dynamic equilibrium regime, that more closely resembles traditional atomic BEC systems.

\begin{methods}

To investigate the coherence time of polaritons in the condensed phase we employ a GaAs based microcavity with InGaAs quantum wells described in our previous work~\cite{cilibrizzi_polariton_2014}. The sample is held at 6K in a cold finger continuous flow cryostat and polaritons are injected non-resonantly with the use of a single mode actively stabilized ultra narrow linewidth ($\approx 100kHz$) CW optical source. We employ a spatial light modulator (SLM) in order to shape the optical beam into a top hat excitation profile or an annular ring. As shown previously~\cite{askitopoulos_polariton_2013}, the ring excitation induces a parabolic potential where polaritons accumulate and upon reaching a threshold density they undergo a phase transition to a BEC. In this configuration the order parameter of the system appears in the form of circular polarization even when the polarization of the excitation is linear ~\cite{askitopoulos_nonresonant_2016,pickup_optical_2018}. However, in order to ensure that the resulting condensate has a dominant population of a single spin component far above threshold, we circularly polarize our excitation allowing us to further filter effects originating from cross spin interactions~\cite{vladimirova_polariton-polariton_2010}.

\subsection{First order Coherence time measurements.}

The first order coherence of the system is obtained by directing the condensate luminescence to an actively stabilized Michelson interferometer equipped with a retro-reflector and a ($\unit[15]{cm}$) double pass delay stage. Experiments were performed at a detuning of $\approx \unit[-4.5]{meV}$ with an exciton Hopfield coefficient of $X=0.2$. The resulting interferograms for the TH and OT condensates are recorded over $\unit[5]{\mu s}$ with a Qimaging Exi camera and the first order spatial coherence $g^{(1)}(x,\tau_i)$ is extracted for different delay times $\tau_i$ (see also SI).
\begin{equation}
g^{(1)}(x,\tau_i)= \frac{ |\langle E^{*}(x,t)E(x,t+\tau_i) \rangle|}{\sqrt{\langle |E(x,t)|^2 \rangle\langle| E(x,t+\tau_i	) |^2\rangle }}
\end{equation}

\end{methods}

\bibliographystyle{naturemag}
\bibliography{coherence}


\begin{addendum}
\item A.A. would like to acknowledge useful discussions with H. Sigurdson, A. Nalitov, J. T{\"o}pfer and S. Tsintzos. The authors acknowledge the support of the Skoltech NGP Program (Skoltech-MIT joint project), and the UK’s Engineering and Physical Sciences Research Council (grant EP/M025330/1 on Hybrid Polaritonics).
\item[Competing Interests] The authors declare no competing financial interests.
\item[Contributions]A.A. and P.L. conceived and designed the research, A.A. and L.P. conducted the first order coherence measurements, L.P. and S.A implemented the active density stabilization, K.L. designed and built the active interferometer stabilization, A.Z. built and characterized the HBT setup and with S.A. and A.A. conducted the second order coherence measurements. A.A. analyzed the data and made the parameter space diagrams,  W.L. and P.L. designed the micro-cavity structure. A.A. and P.L. wrote the paper with input from all authors.
\item[Correspondence] Correspondence and requests for materials should be addressed to \newline A.A. ~(email : A.Askitopoulos@skoltech.ru).
\end{addendum}

\end{document}


\title{Supplementary Information for Giant increase of temporal Coherence in a polariton condensate}

\date{\today}
\author{A. Askitopoulos}
\affiliation{Skolkovo Institute of Science and Technology Novaya St., 100, Skolkovo 143025, Russian Federation}
\author{L. Pickup}
\affiliation{School of Physics and Astronomy, University of Southampton, Southampton, SO171BJ, United Kingdom}
\author{S.Alyatkin}
\affiliation{Skolkovo Institute of Science and Technology Novaya St., 100, Skolkovo 143025, Russian Federation}
\author{A. Zasedatelev}
\affiliation{Skolkovo Institute of Science and Technology Novaya St., 100, Skolkovo 143025, Russian Federation}
\author{K.G.Lagoudakis}
\affiliation{Department of Physics, University of Strathclyde, Glasgow, G40NG,United Kingdom}
\author{W. Langbein}
\affiliation{School of Physics and Astronomy, Cardiff University, Cardiff CF243AA, United Kingdom}
\author{P.G. Lagoudakis}
\affiliation{Skolkovo Institute of Science and Technology Novaya St., 100, Skolkovo 143025, Russian Federation}
\affiliation{School of Physics and Astronomy, University of Southampton, Southampton, SO171BJ, United Kingdom}
\maketitle


\section{Experimental Configuration}\label{SupES}

\begin{figure}[ht]
	\center
	\includegraphics[scale=0.35]{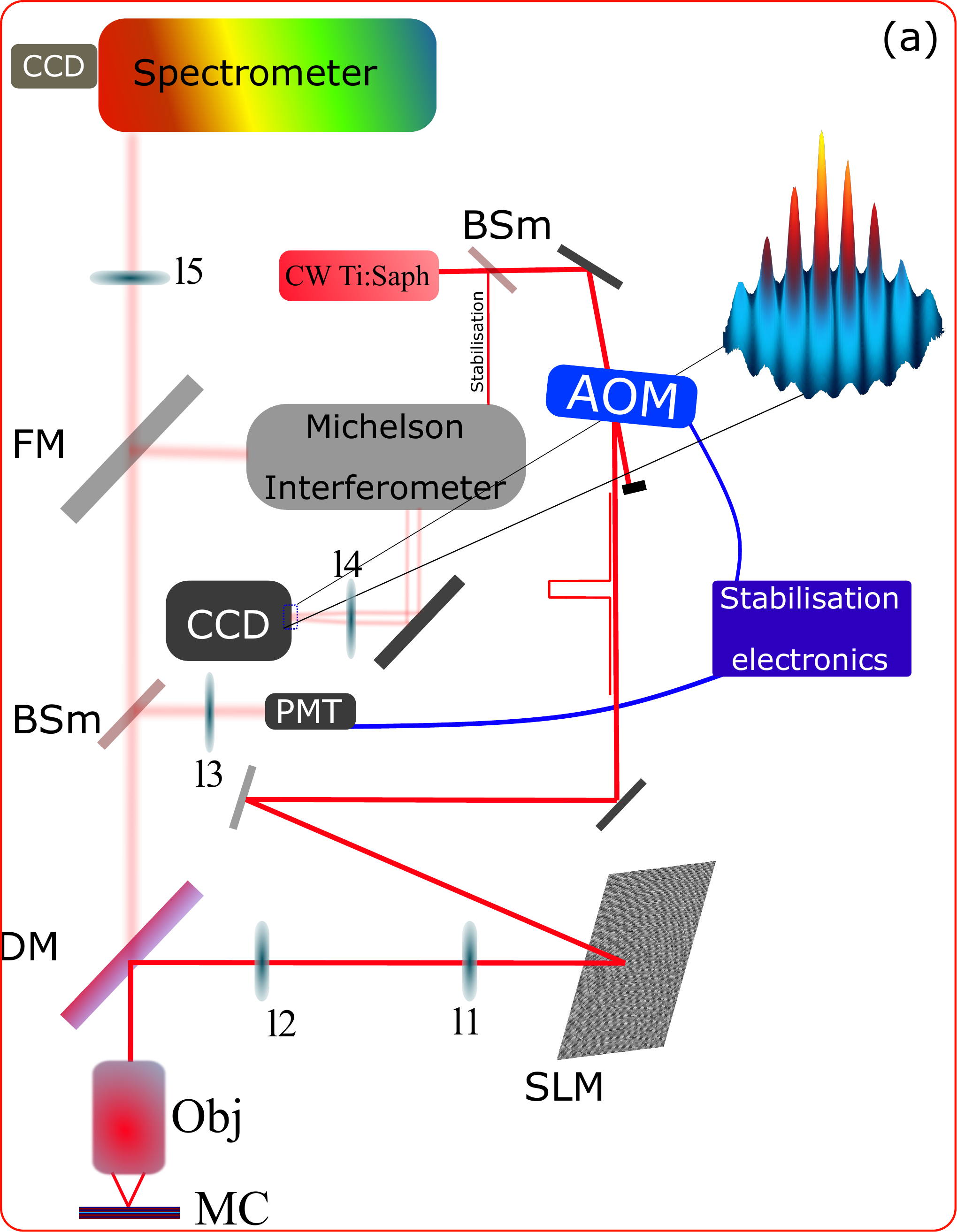}
	\centering
	\caption{Schematic representation of the experimental setup and typical interference fringe pattern recorded plotted in 3d. l1-l5 relay lenses, BS beam splitter, BSm 1\% beam sampler, DM dichroic mirror, Obj Objective lens, FM  flipper mirror, MC microcavity sample, PMT Photomultiplier tube and SLM spatial light modulator with a representative phase pattern. }
	\label{sfig1}
\end{figure}

The optical excitation is modulated with the use of an acousto-optic modulator (AOM), we create $50\mu s$ quasi-pulses with less than $1\%$ duty cycle and project them on the sample with a confocal objective of NA=0.4. The emission from the sample is collected with the same objective and sent either to a varying delay Michelson interferometer, actively stabilized with the same laser with the use of a piezo and a PID and then projected to a charged coupled device (CCD) or to a $\unit[750]{mm}$ spectrometer equipped with a CCD \fig{sfig1}. To further avoid any instabilities that will degrade the interference pattern, we use single shot triggered acquisition of $5\mu s$. The condensate density is actively stabilized against intensity fluctuations with a photomultiplier tube (PMT) that provides feedback to the AOM signal (see section Density stabilization ).
By scanning the delay line of our interferometer we can extract the decay of the first order coherence function $g^{(1)}[x,y,\tau$]. Although performing a phase scan acquisition for every delay step would be the most accurate measurement, we can nevertheless extract the information about the decay of first order coherence by acquiring an interferogram for each delay point along with individual measurements of the two reference arms.

\section{Analysis of first order coherence}
For the first order coherence time measurements, the real space intensity distribution $I_{if}(x,y)$ measured by our camera can be represented by

\begin{align}
I_{if}(x,y) & =I_{1}(x,y)+I_{2}(x,y)+|g^{(1)}(x,y,\tau)|2 \sqrt{I_{1}(x,y)I_{2}(x,y)}cos(k r + \phi)\label{Seq1}  \\
& = I_{1+2}\left( 1+ \eta|g^{(1)}(r,\tau)|cos(k r + \phi) \right) \label{Seq2}
\end{align}

Where $I_1$ and $I_2$ are the individual intensities of the two arms, $I_{1+2}(x,y)=(I_{1}(x,y)+I_{2}(x,y)) $ ,  $I_{12}(x,y)=2\sqrt{I_{1}(x,y)I_{2}(x,y)} $ and $\eta={I_{12}(x,y)}/{I_{1+2}(x,y)}$ and $I_{if}(x,y)$ is the interferogram recorded by the camera. For a Gaussian intensity profile of $I_{1,2}$ we fit the one dimensional profiles from the interferogram image, integrated around different points of the y axis with equation 2 and extract  $\tilde{g}^{(1)}(x,\tau)=\eta*\left|g^{(1)}(x,\tau)\right|$ (see also supplementary video for an animation of the fitting procedure of a single interferogram). We then scan the 15$\unit{cm}$ double pass delay stage and record the interferograms for different delays. For every delay step, the reference intensities $I_{1},I_{2}$ are also recorded. For perfectly balanced intensities $\eta=1$, however due to experimental limitations (e.g. the reflection efficiency of the retro is lower then the mirror), we use the averaged reference intensities to extract $\eta$ and normalize the extracted $\tilde{g}^{(1)}(x,\tau)$. This process is repeated for increasing polariton densities for 2 detunings, while the excitation geometry is left unaltered. All data were recorded for individual optical excitation pulses of 20-50$\unit{\mu sec}$ within an integration time of $5 \unit{\mu s}$, while the shot to shot delay was of the order of $\unit[1]{ms}$ with an effective duty cycle of less than $\unit[0.5]{\%}$.



\section{Coherence times for different Interaction Strengths}
The polariton-polariton interaction strength ($U_{pp}$) has been the subject of a rigorous scientific debate as its reported values, measured by means of the polariton energy shift with density, spans over 4 orders of magnitude ~\cite{tassone_exciton-exciton_1999,love_intrinsic_2008,sun_direct_2017,tsintzos_electrical_2018,vladimirova_polarization_2009,estrecho_direct_2019}, although initial theoretical calculations~\cite{tassone_exciton-exciton_1999} have been successfully used for the modelling of the condensate properties giving very good agreement with experimental findings. However, as the exciton polariton interaction, also contributes to the condensate energy shift, recent studies have measured this shift in the OT configuration where the condensate is spatially separated from the pump induced reservoir interactions~\cite{sun_direct_2017,estrecho_direct_2019}. Nevertheless, the trap ground state energy is also affected from the change in reservoir density that effectively changes the trap potential~\cite{askitopoulos_robust_2015} and this needs to also be accounted for (see also SI). In this regime a measurement of the coherence time can unambiguously settle the debate on the interaction strength of polaritons as it can be used to estimate the upper bound of the interaction strength~\cite{love_intrinsic_2008}, under the assumption of a fully second order coherent state. For an exciton interaction strength of $U_{xx}=\unit[2]{\mu eV \mu m^2}$~\cite{tassone_exciton-exciton_1999}, a polariton excitonic fraction (X) of 0.1 and a condensate density of $n_c=\unit[200]{\mu m^{-2}}$ and disregarding Schawlow-Townes effects, a polariton condensate obeying Poissonian Statistics, would have a coherence time $t_{coh}\approxeq\unit[1]{ns}$. For the highest literature value of $U_{xx}=\unit[1.74]{meV \mu m^2}$~\cite{sun_direct_2017}, the same condensate would have a coherence time of only $t_{coh}\approxeq\unit[1.2]{ps}$.
%

\section{Condensate density}\label{ConD}
The condensation density was evaluated with three distinct methods. Firstly, by taking into account the detector quantum efficiency for the condensate energy, the double sided emission of the microcavity sample, the optical path efficiency, the photon to electron conversion rate of the camera, the number of counts on the camera integrated from inside an area with $I_{pix}\geq0.1I_{max}$, the exposure time and using a polariton lifetime of ~\unit[7]{ps}. All the polariton population values reported in the manuscript were calculated through this method and for $P\approxeq P_{th}$, $\bar{n}_{th}=656$. Using eq.1 of the main text and  the experimentally measured values of $g^{(2)}(\tau=0,\bar{n}_{th})-1=1.079$ and $\tau_{g2}(\bar{n}_{th})=\unit[400]{ps}$ and polariton lifetime $t_p=\unit[7]{ps}$, we get $\bar{n}_{th}=\frac{\tau_{g(2)}(\bar{n}_{th})-t_p}{t_p(g^{(2)}(\tau=0,\bar{n}_{th})-1)}=703$ in very good agreement with the previous method. Finally we measure the output optical power of the condensate at threshold taking into account again the detection efficiency and the double sided emission and we find $n_{th}\approxeq250\pm30$ which is of the same magnitude of the previous measurements but reflects the limited accuracy of our detector at low powers.

\section{Density stabilization}\label{SSDS}
To achieve density stability across the many condensate realizations in each $g^{(1)}(\tau)$ decay curve, it is necessary to use an additional feedback system that stabilizes the global intensity of the emitted PL. The stabilization system implemented incorporates a photo multiplying tube, the signal of which is sent a PID loop. The PID loop compares the signal to an external reference, that has the same modulation shape and frequency as the excitation beam, resulting in a compensation signal that is subsequently summed with the voltage train driving the AOM. Due to the AOM modulation frequency and duty cycle, it is not possible to stabilize within a singular CW pulse. However, the stabilization system works to make the integrated photoluminescence intensity stable, at the desired level, across many CW pulses. This system compensates for example, small fluctuations and drifts in laser intensity or remaining vibrations on the cryostat, which can result in significant density fluctuations due to the strong non-linear behaviour of the intensity around threshold. This also allows to keep the excitation density very accurately around the threshold for the second order coherence measurements and contributes to the clear bunching signal that was observed.

\section{Photon correlation measurements}

Second-order time correlations of the polariton emission are investigated under the same excitation conditions using a heralded Hanbury Brown and Twiss (HBT) configuration~\cite{brown_correlation_1956}. The light is split in two detection channels by a 50:50 fiber beam splitter (Newport F-CPL-M22855-FCPC), and the single photon detection events recorded with single-photon Avalanche photo diodes (SPAPD) (IDQ 100). One channel is used as a start trigger and the other acts as a stop signal for a TCSPC module (SPC-160, Becker \& Hickl GMBH). Detection coincidence events are counted within a 12 ps width of a single time channel. The APDs have a time resolution as short as 60 ps and a dark count rate of 40 Hz. The overall time resolution of the setup is 100 ps for counting rates lower than $10^6 s ^{-1}$. The photon counting rate was kept below $10^5 s^{-1}$ in all the experiments. Second-order correlation measurements of the condensate were carried out in the OT and TH configurations. The excitation densities are smoothly altered around the threshold optical power density for both configurations.
For the number of detected photons N in each of the detection channels (for Poissonian statistics) the uncertainty scales as $\sqrt{N}$ . Therefore, the signal to noise ratio scales as square root of N and consequently proportional to square root of total integration time.
\section{Saturation density}
\begin{figure}[ht]
	\center
	\includegraphics[scale=0.65]{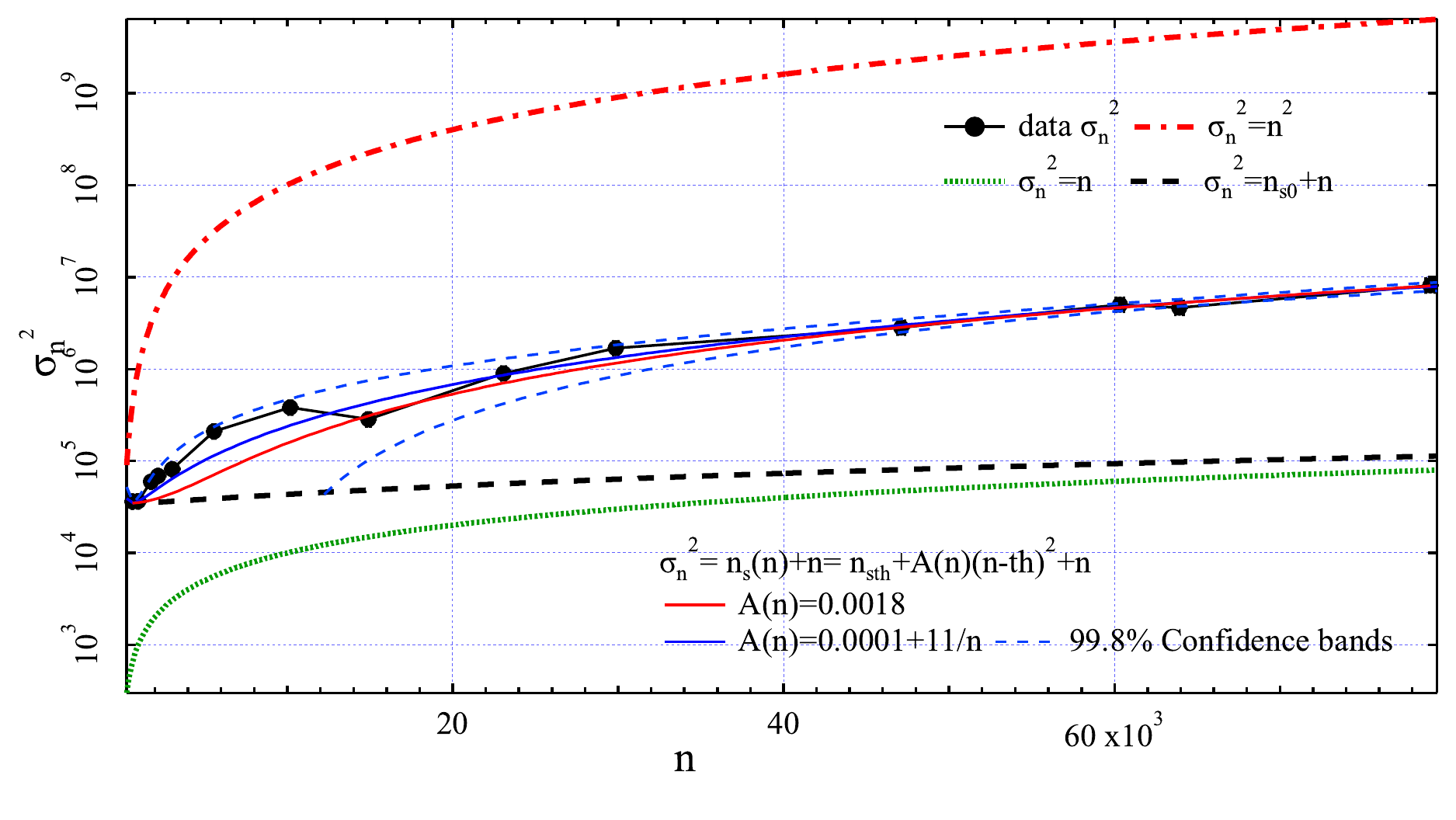}
	\centering
	\caption{Polariton density variance $\sigma_{n}^2$ as extracted from the ensemble of measurements, black circles, along with a fit to the data (solid blue line) and confidence interval prediction bands (dashed blue lines). The black dashed line represents the case where $n_s(n)=n_{s0}$ while green dotted and red dashed-dotted denote coherent and chaotic-thermal statistics respectively.}
	\label{sfig3}
\end{figure}
The term $n_s$ in single mode quantum laser theory is refereed to as the "saturation photon number"~\cite{loudon_quantum_2000}, and is defined as 
\begin{align} 
&n_s=\frac{\Gamma_{sp}}{\Gamma_{st}} \\
\Gamma_{sp}=2\gamma_{sp}  
&	\hspace{2cm} \Gamma_{st}=\frac{2\pi c^3 \gamma_{sp}}{V\omega_0^2 \gamma}
\label{Ns}
\end{align}
where $\gamma_{sp}$ is the spontaneous emission rate, $\omega_0$ and $\gamma$ are the atomic transition energy and linewidth respectively and V is the cavity volume. Effectively the ratio $\nicefrac{1}{n_s}=\beta$ quantifies the fraction of the spontaneous emission into the lasing mode~\cite{rice_photon_1994} and for $\beta=1$ we have an apparently threshold-less laser~\cite{yokoyama_physics_1992}. 
For polaritons, we can approximate the spontaneous contribution to the condensate mode with the thermalization processes that cause polaritons to relax and redistribute their population in the lower polariton dispersion. Additionally, as polariton bosonic stimulation manifests as stimulated scattering from a high energy and concentration polariton state to the condensate energy, the energy difference defined by $\omega_0$ will correspond to the initial and final states of the stimulation process. For excitation geometries that result in condensates formed at the bottom of the dispersion at $k_{\parallel}=0$, as is the case for the ground state condensation in the optical trap and top hat excitation sizes of $r>10$,  this term will increase for more negatively detuned systems.

Regarding the contribution of $n_s$ to the particle statistics of the system, as mentioned in the main text, $\sigma_n^2=n_s+n$ which for the case of a conventional laser with $n_s=n_{s0}$, the variance increases linearly with the mode population (dashed black line ~\fig{sfig3}). The recorded data (~\fig{sfig3} black circles) for the variance of the system demonstrate that $\sigma_n^2$, increases faster with density which can only arise due to an increase of $n_s$. The boundaries between which $n_s=f(n)$ evolves will be the limiting case of a fully coherent mode, $\sigma_n^2=\bar{n}$, and one with thermal/chaotic statistics $\sigma_n^2=\bar{n}$. It is therefore reasonable to a assume that above threshold the additional particles that populate the condensate will have a probability of contributing with either $n^2$ or n to the total fluctuations. An equivalent way of describing this effect is that as the energy of the final state of the stimulation process changes, the threshold density for condensation is also affected, leading to a change in the number of particles in the mode that originate from spontaneous relaxation.
Effectively this translates to
\begin{equation}
 \sigma_{\bar{n}}^2(\bar{n})= n_{s}(\bar{n}) +\bar{n}= n_{s}(\bar{n}_{th})+ A(n)(\bar{n}-\bar{n}_{th})^2+\bar{n}
 \label{eqns}
\end{equation}
For the lasing phase transition $A(\bar{n})=0$, and the gain saturation is not affected by the number of particles in the mode as there are no interactions that change the energy levels of the stimulation process. In polariton systems on the other hand the interactions between particles can shift the energy level of initial and final state changing the effective threshold of the system with density. This is a distinct difference in the above threshold statistics of polariton systems, where the particles can interact and scatter with each other, and conventional inversion based lasers. It also demonstrates that this difference effectively scales and becomes more pronounced with the interaction strength of polaritons (i.e. the more excitonic or matter based the system becomes) while in the limit of zero interaction it reverts to a normal lasing system. In our analysis this entire dependence is masked in the parameter $A(\bar{n})$ which we have approximated in two different ways. Firstly with a simple linear dependence on $(\bar{n}-\bar{n_th})^2$), red continuous line in \fig{sfig3}, which corresponds to a linear shift with energy above threshold. However, as the density is increased and the interactions are gradually quenched we approximate also with a sub-linear dependence which is shown with blue continuous line in \fig{sfig3}. In general, a microscopic theory for the exact form of this dependence should be developed but this is beyond the scope of this work.


\section{Condensate Size and linear properties of the optical trap}

It is worth examining the trapped wave-function properties in respect to the geometric parameters of the potential to elucidate the expected dynamic behavior of the system and examine the correspondence with the experimental results. For this we use a linear time independent 1D quantum model of the system taking into account the values of the experimental polariton mass to simulate the steady state of the optical trap and extract the ground state wave-function in the single particle regime (absent of any nonlinear interactions). We use a finite real trapping potential composed of two Gaussian profiles and employ the Numerov method~\cite{kalogiratou_numerical_2005} to find the bounded eigen-energies and eigen-functions of  the potential.
\begin{figure}[ht]
	\center
	\includegraphics[scale=0.45]{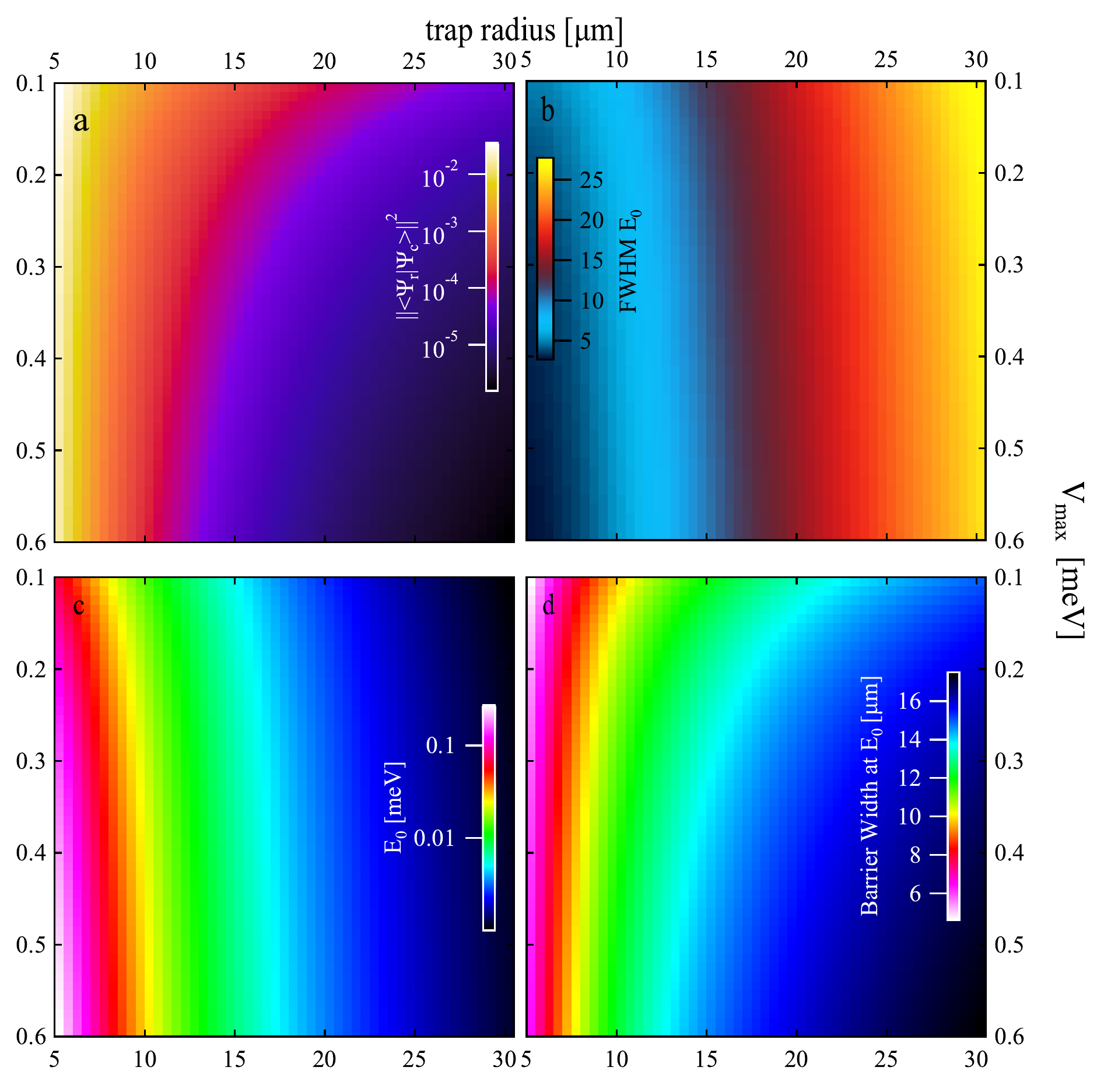}
	\centering
	\vspace{-10pt}
	\caption{Quantitative analysis of the 1D optical trap for varying potential depth and trap radius. (a) Calculated overlap of reservoir ($\Psi_r$) and ground state mode ($\Psi_c$). (b) Calculated spatial extent (FWHM) of the trap ground state. Dependence of the ground state energy $E_0$ (c) and of the barrier width at the ground state energy level. The FWHM of the Gaussian barrier is $\unit[3.5]{ \mu m}$.}
	\label{sfig4}
	\vspace{-10pt}
\end{figure}

We first examine the parameter space of potential depth and trap radius(\fig{sfig4}). As expected, the reservoir-wavefunction probability overlap integral that effectively showcases the strength of the condensate interaction with the reservoir, diminishes with trap size and depth and although it is non-vanishing even for traps with radii of $r_c=\unit[30]{\mu m}$, it nevertheless is a very small fraction, much larger than the $\nicefrac{U_{xp}}{U_{pp}}$ ratio. Additionally the energy shifting of the ground state for increasing trap depth, which relates to increased pumping of the reservoir, is demonstrated. Indeed, as we argue in the main text this energy shifting, depending on the trap parameters can be comparable to the one induced by repulsive polariton interactions.
Lastly, as exemplified by the dependence of the barrier width,  we observe that the mode confinement is strengthened as the trap depth, which relates to the density of particles in the reservoir, is increased and this also translates to reduced overlap with the reservoir.
\begin{figure}[ht]
	\center
	\includegraphics[scale=0.45]{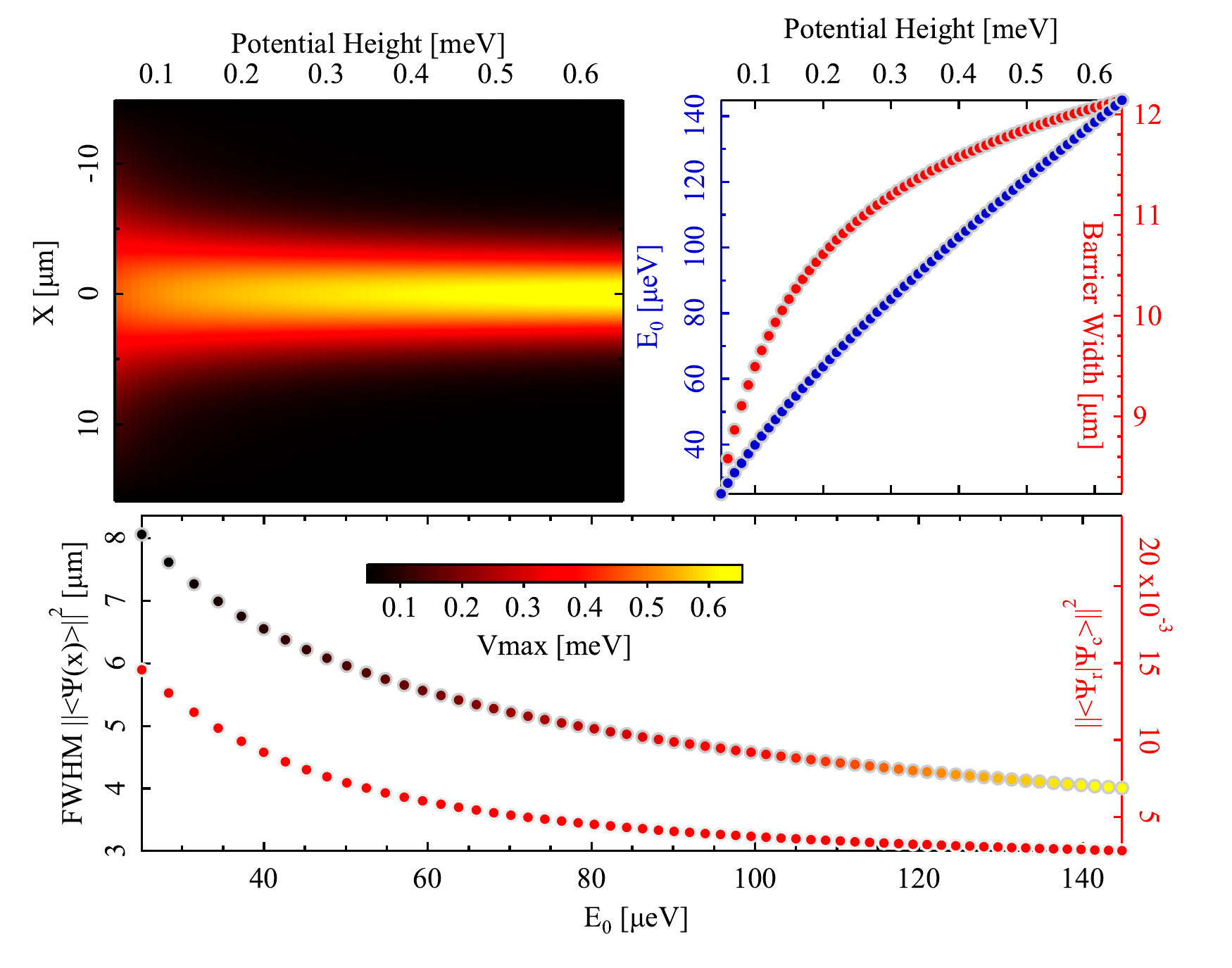}
	\centering
	\vspace{-10pt}
	\caption{Wavefunction evolution for a 1D potential with values close to the experimental ones. (a) evolution of the 1D profile of $\Psi_0(x)$ for increasing trap depth. (b) Energy $E_0$ and corresponding barrier width at $E_0$ vs potential height. (c) FWHM of $\Psi_0(x)$ and wavefunction reservoir probability overlap integral vs energy, the FWHM is colorcoded with the corresponding potential height. }
	\vspace{-10pt}
	\label{sfig6}
\end{figure}

\begin{figure}[ht]
	\center
	\includegraphics[scale=0.45]{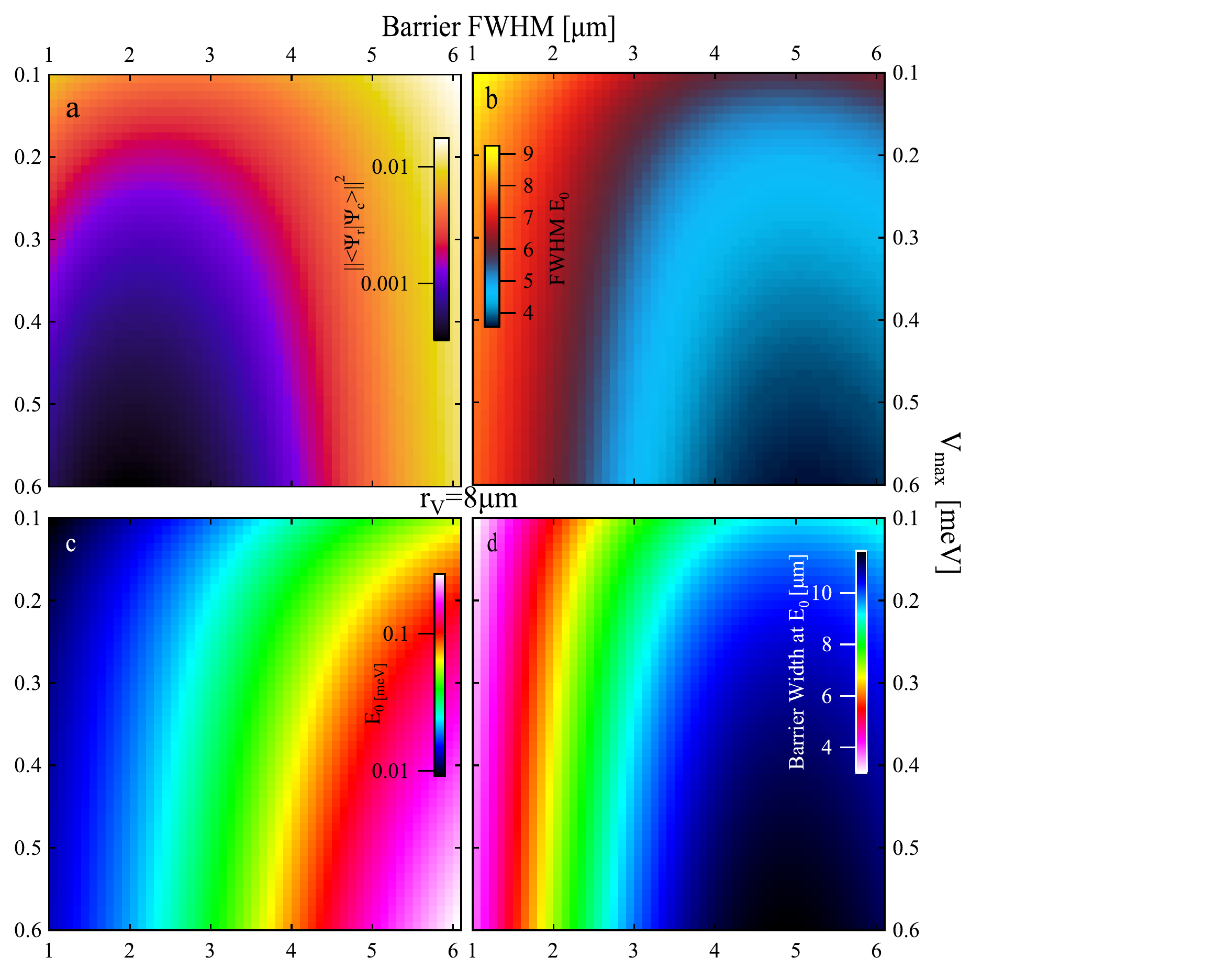}
	\centering
	\vspace{-10pt}
	\caption{Quantitative analysis of the 1D optical trap for varying potential depth and barrier FWHM. (a) Calculated probability overlap integral of reservoir ($\Psi_r$) and ground state mode ($\Psi_c$). (b) Calculated spatial extent (FWHM) of the trap ground state. Dependence of the ground state energy $E_0$ (c) and of the barrier width at the ground state energy level (d)}
	\vspace{-10pt}
	\label{sfig5}
\end{figure}
In the experimental study the optical trap was initiated with a radius of $r_{OT}=\unit[8.5]{\mu m}$ with a barrier FWHM of $\approx \unit[3.5]{\mu m}$ and this regime for varying potential height is shown in~\fig{sfig6}. While the initial FWHM of the wavefunction is quite similar to the experimental one ($FWHM_{exp}=\unit[6.6]{\mu m}$) interestingly, for increasing densities and hence potential height the mode is focused as the effects of confinement become more pronounced, in contrast to the experimental behaviour were the condensate FWHM expands around 30\% within the power range studied. Nevertheless, this in fact demonstrates the presence of the repulsive polariton-polariton interactions that tend to counter balance this effect. Indeed, the condensate size dependence on density can also be used to estimate the interparticle interactions in the condensate. More importantly, we note that the condensate reservoir overlap for this 1D model is of the order of $1\%$. For a 2d system this value is expected to be even smaller, for instance for a 1D potential  and wave-function with a probability overlap integral of $|\left\langle\Psi_r|\Psi_{0} \right\rangle|^2_{1D}\approxeq0.01$, the corresponding 2 dimensional wave-functions have a probability overlap integral  $|\left\langle\Psi_r|\Psi_{0} \right\rangle|^2_{2D}\approxeq10^{-5}$ or a ratio of $\nicefrac{|\left\langle\Psi_r|\Psi_{0} \right\rangle|^2_{1D}}{|\left\langle\Psi_r|\Psi_{0} \right\rangle|^2_{2D}}= 985 $. Furthermore, this first approximation of the ground state wave function doesn't account for dynamic effects like the increased depletion of the reservoir underneath the wave-function that will in turn reshape reservoir and ground state further diminishing the overlap integral~\cite{estrecho_single-shot_2018}. Lastly, taking also into account that the condensate density $N_c$, above threshold, is greater than the density of the un-condensed reservoir, it is evident that exciton-polariton interactions will have a marginal effect on the dephasing of the condensate. 


To conclude, we complement our examination of the parameter space of the trapping potential by performing an analysis for the ground state, but for varying FWHM of the potential barrier for a fixed radius of $r=\unit[8]{\mu m}$ and this is shown in \fig{sfig5}. 

\section{Intensity Noise fluctuations}
\label{INF}

\begin{figure}[ht]
	\center
	\includegraphics[scale=0.61]{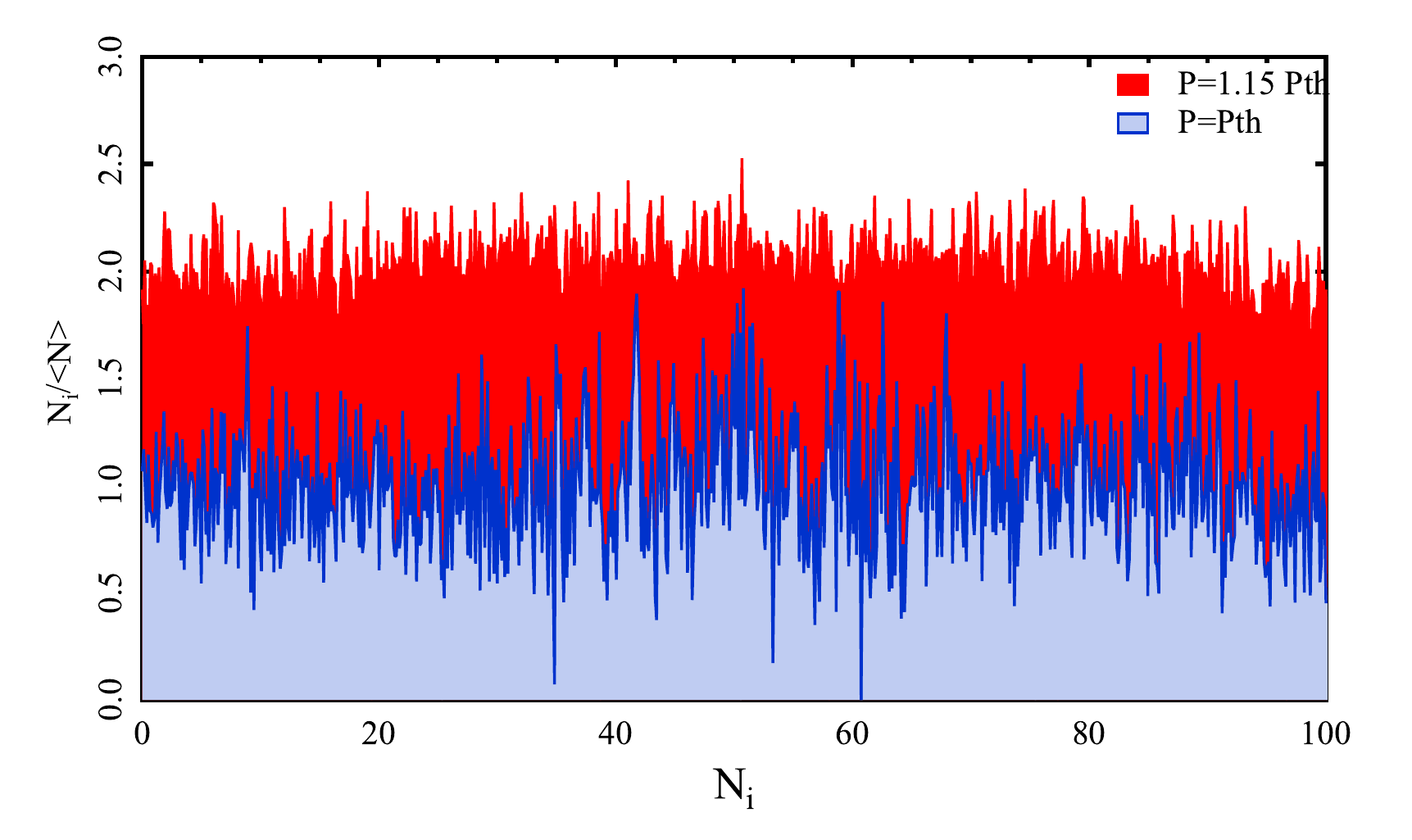}
	\centering
	\caption{Density fluctuations of the condensate for individual realizations as recorded within the $\unit[5]{\mu s}$ integration window normalized on the average density for $P=P_{th}$ blue and $P=1.15P_{th}$ red. The red curve has been shifted vertically by 1 for clarity. The horizontal axis is in \% of measured realizations.}
	\label{sfig2}
\end{figure}

To extract the variance ($\sigma_{n}^2$) of the condensate density for increasing powers above threshold, we use the reference images recorded for a single arm of the interferometer. For each delay scan, the camera records the reference images for every delay step ($i$) leading to a statistically significant ensemble ($250<i<1500$). For each acquisition, we normalize the counts for the detection efficiency of our detection path, the integration time and the polariton lifetime ($t_p=\unit[7]{ps}$). We then fit a 2-dimensional Gaussian to the data and extract the total number of counts under the curve that corresponds to the total polaritons in the mode for this realization. From the ensemble of realizations that we have recorded, we extract the variance and average density of polaritons in the condensate mode. Figure \fig{sfig2} shows the shot to shot intensity of two typical powers above threshold, normalized with the average intensity, where the fluctuations for the threshold power density is clearly higher.

We note that in order for this approach to be valid the shot to shot signal noise should be greater than the shot to shot readout noise of the camera. For all the data presented here the single image, single pixel $4<SNR<11$ while the readout noise of the area of interest of the camera is $\sigma_{Ib}^2=34.08$.
A similar method has previously been used for noise estimation even beyond the shot noise limit for atomic condensates~\cite{riedel_atom-chip-based_2010}. 

\begin{figure}[ht]
	\center
	\includegraphics[scale=0.4]{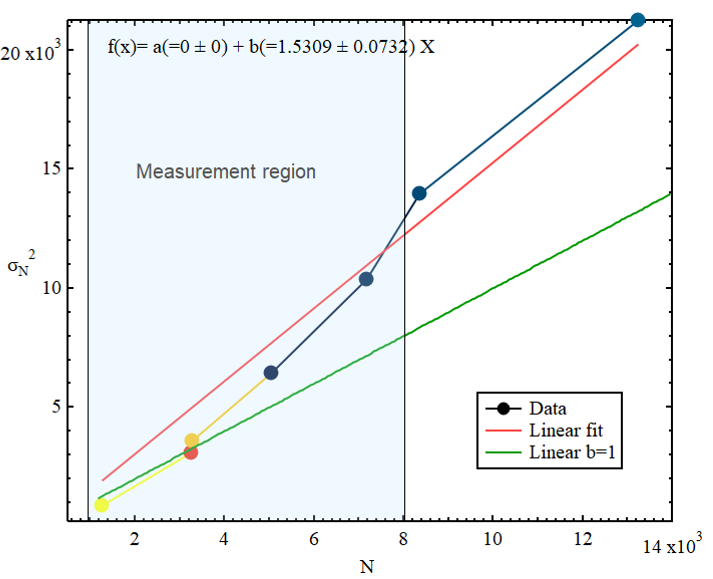}
	\centering
	\caption{Recorded Laser variance $\sigma_{n}^2$ as a function of photoelectrons per pixel. The noise of the system is almost linear and very close to pure Poissonian statistics} 
	\label{sfig7}
\end{figure}
We also evaluate the linearity of the noise of our imaging apparatus for varying signal to noise ratio, by conducting a reference measurement of the noise of our stabilised CW optical source for a range of optical densities on the detector. The results are displayed in ~fig.\fig{sfig7} where we observe that the dependence of the measured noise on the optical signal on our detector is approximately linear and indeed very  close to theoretical values. The linearity of the noise demonstrates, similarly to previous experiments with atomic condensates that the signal noise dominates over readout and electronic noise of the system as well as the classical noise of the laser~\cite{riedel_atom-chip-based_2010,iskhakov_intensity_2011} and  cannot adequately account for the excess super-Poissonian statistics of the condensate displayed in \fig{sfig3}.

\section{Supplementary Videos}

\begin{itemize}
	\item Video S1: Characteristic fitting for profiles of a single interferogram $I_{if}$ along the lateral direction of the interference fringes with use of ~eq.\ref{Seq2} and extraction of $g^{(1)}(y,\tau_i)$.
	\item Video S2: Decay of $g^{(1)}(\tau,n)$ for different densities n of the condensate in the optical trap,  together with Gaussian fits and fitting prediction bands, left panel and extracted $t_{coh}$ versus density ,right panel
	\item Video S3: Evaluation of coherence time from equation 2 of the main text, versus polariton density and polariton interaction strength for different values of gain saturation ($n_s$) with logarithmic colorscale and profiles corresponding to $U_{pp}=0.01\mu eV$, blue line, and to polariton density $N_p=1000$, red line.
\end{itemize}
%